\documentclass[12pt,pre,aps,preprint,showpacs]{revtex4}

\usepackage{epsfig}
\usepackage{graphicx}
\usepackage{amsfonts}
\usepackage{amsmath}
\usepackage{float}
\usepackage{multirow}

\begin{document}

\title{Precise Algorithm to Generate Random Sequential Addition of Hard Hyperspheres at Saturation}

\author{G. Zhang}

\email{gezhang@princeton.edu}

\affiliation{\emph{Department of Chemistry}, \emph{Princeton University},
Princeton NJ 08544}

\author{S. Torquato}

\email{torquato@electron.princeton.edu}

\affiliation{\emph{Department of Chemistry, Department of Physics,
Princeton Institute for the Science and Technology of
Materials, and Program in Applied and Computational Mathematics}, \emph{Princeton University},
Princeton NJ 08544}


\begin{abstract}
The study of the packing of hard hyperspheres in $d$-dimensional Euclidean space $\mathbb{R}^d$ 
has been a topic of great interest in statistical mechanics and condensed matter theory. 
While the densest known packings are ordered in sufficiently low dimensions, it has been suggested that in sufficiently large dimensions, the densest packings might be disordered. 
Random sequential addition (RSA) time-dependent packing process, 
in which congruent hard hyperspheres are randomly and sequentially placed into a system without interparticle overlap, 
is a useful packing model to study disorder in high dimensions. 
Of particular interest is the infinite-time {\it saturation}
limit in which the available space for another sphere tends to zero. 
However, the associated saturation density has been determined in all previous investigations by extrapolating
the density results for near-saturation configurations to the saturation limit, which necessarily introduces numerical uncertainties. 
We have refined an algorithm devised by us [S. Torquato, O. Uche, and F.~H. Stillinger, Phys. Rev. E {\bf 74},  061308 (2006)] to generate RSA packings
of identical hyperspheres.
The improved algorithm produce such packings that are guaranteed to contain no available space using finite computational time 
with heretofore unattained precision and across the widest range of dimensions ($2 \le d \le 8$).
We have also calculated the packing and covering densities, pair correlation function $g_2(r)$ and structure factor $S(k)$ of the saturated RSA configurations.
As the space dimension increases, we find that pair correlations markedly diminish, consistent with a recently proposed ``decorrelation'' principle,
and the degree of ``hyperuniformity" (suppression of infinite-wavelength density fluctuations) increases. 
We have also calculated the void exclusion probability in order to compute the so-called quantizer error of the RSA packings, 
which is related to the second moment of inertia of the average Voronoi cell.
Our algorithm is easily generalizable to generate saturated RSA packings of nonspherical particles.
\end{abstract}

\maketitle
\section{Introduction}
\label{Introduction}

In $d$-dimensional Euclidean space $\mathbb{R}^d$, a hard hypersphere (i.e. $d$-dimensional sphere) packing is an arrangement of hyperspheres in which no two hyperspheres overlap. 
The \emph{packing density} or \emph{packing fraction} $\phi$ is the fraction of space in $\mathbb{R}^d$ covered by the spheres, 
which for identical spheres of radius $R$, the focus of the paper, is given by:
\begin{equation}
\phi=\rho v_1(R),
\end{equation}
where $\rho$ is the number density and
\begin{equation}
v_1(R)=\frac{\pi^{d/2}}{\Gamma(1+d/2)}R^d
\end{equation}
is the volume of a $d$-dimensional sphere of radius $R$ and $\Gamma(x)$ is the gamma function.
Sphere packings are of importance in a variety of contexts in the physical and mathematical sciences.
Dense sphere packings have been used to model a variety of many-particle systems, including liquids \cite{hansen_2006_liquid}, 
amorphous materials and glassy states of matter \cite{bernal_1960_geometry, frisch_1999_glass,Parisi_2000_glass,torquato_2002_controlling,parisi_2006_amorphous, parisi2010mean, torquato2010jammed}, 
granular media \cite{torquato_2001_random}, suspensions and composites \cite{chan1991effective, zohdi2006optical, mejdoubi2007numerical}, and crystals \cite{chaikin_2000_principles}. 
The densest sphere packings are
intimately related to the ground
states of matter \cite{chaikin_2000_principles, torquato_2010_reformulation} and the optimal way of sending digital signals over noisy
channels \cite{conway_1998_packing}. Finding the densest sphere
packing in $\mathbb{R}^d$ for $d\ge 3$ is generally a notoriously
difficult problem \cite{conway_1998_packing}. Kepler's conjecture, 
which states that there is no other three-dimensional arrangement of identical spheres 
with a density greater than that of face-centered cubic lattice, 
was only recently proved \cite{hales_2005_kepler_conjecture_proof}. 
The densest sphere packing problem in the case of congruent spheres has not been rigorously solved for $d \ge 4$ \cite{conway_1998_packing, cohn_2003_upper_bound_packing}, 
although for $d=8$ and $d=24$ the $E_8$ and Leech lattices, respectively,
are almost surely the optimal solutions \cite{cohn_2004_optimality}.

Understanding the high-dimensional behavior of disordered sphere packings is a fundamentally important problem,
especially in light of the recent conjecture that the densest packings in sufficiently high dimensions may be disordered rather than ordered \cite{torquato_2006_lower_bound_packing}. 
Indeed, Ref.~\onlinecite{torquato_2006_lower_bound_packing} provides a putative exponential improvement
on Minkowski's lower bound on the maximal density $\phi_{max}$ among all Bravais lattices \cite{minkowski_1905_bound}:
\begin{equation}
\phi_{max} \ge \frac{\zeta(d)}{2^{d-1}},
\label{minkowski_bound}
\end{equation}
where $\zeta(d)=\sum_{k=1}^{\infty} k^{-d}$ is the Riemann zeta function. For large values of $d$, the asymptotic behavior of the Minkowski's lower bound is controlled by $2^{-d}$. 
Interestingly, any saturated packing density satisfies the following so-called ``greedy'' lower bound:
\begin{equation}
\phi \ge \frac{1}{2^d}.
\label{greedy_bound}
\end{equation}
A saturated packing of congruent spheres of unit diameter and density $\phi$ in $\mathbb{R}^d$ 
has the property that each point in space lies within a unit distance from the center of some sphere. 
Thus, a covering of the space is achieved if each center is encompassed by a sphere of unit radius and the density of this covering is
\begin{equation}
\theta=2^d\phi \ge 1,
\label{covering_density_formula}
\end{equation}
which proves the lower bound (\ref{greedy_bound}). 
Note that it has the same dominant exponential term as in inequality (\ref{minkowski_bound}). 
The packing density of $2^{-d}$ can also be exactly achieved by {\it ghost random sequential addition} packings \cite{torquato_2006_ghostRSA}, 
an unsaturated packing less dense than the standard random sequential addition (RSA) packing \cite{widom_1966_RSA} in some fixed dimension $d$, 
implying that the latter will have a superior dimensional scaling. 
Additionally, the effect of dimensionality on the behavior
of equilibrium hard-sphere liquids \cite{finken2001freezing, skoge2006packing, rohrmann2007structure, van2009geometrical, lue2010fluid}
and of maximally random jammed spheres \cite{skoge2006packing, parisi2010mean, torquato2010jammed} have been investigated.

Sphere packings are linked to a variety of fundamental
characteristics of point configurations in $\mathbb{R}^d$, including 
the {\it covering radius} $R_c$ and the {\it quantizer error} $\mathcal G$, which are related to properties
of the underlying Voronoi cells \cite{conway_1998_packing}.
The covering and quantizer problems have relevance in numerous applications, including wireless communication
network layouts, the search of high-dimensional data parameter spaces, stereotactic radiation therapy,
data compression, digital communications, meshing of space for numerical analysis, coding, and cryptography \cite{torquato_2010_reformulation, conway_1998_packing}. It has recently been shown \cite{torquato_2010_reformulation} that both of these quantities can be extracted from the {\it void exclusion probability} $E_V(R)$, which is defined to be the probability of finding a randomly placed spherical cavity of radius $R$ empty of any points. It immediately follows that $E_V(R)$ is the expected fraction of space not covered by circumscribing spheres of radius $R$ centered at each point. Thus, if $E_V(R)$ is identically zero for $R \ge R_c$ for a point process, then there is a covering associated
with the point process with covering radius
$R_c$. Finally, for a point configuration with positions
${\bf r}_1,{\bf r}_2, \ldots$, a quantizer is a device that takes as an input a position $\mathbf x$ in $\mathbb R^d$ and outputs the nearest point $\mathbf r_i$ of the configuration to $\mathbf x$. Assuming $\mathbf x$ is uniformly distributed, one can define a mean square error, called the {\it scaled dimensionless quantizer error}, which can be obtained from the void exclusion probability via the relation \cite{torquato_2010_reformulation}:
\begin{equation}
\label{QuantizerDefinition}
\mathcal{G}=\frac{2\rho^{\frac{2}{d}}}{d}\int _0 ^{\infty} R E_V(R) dR.
\end{equation}
It is noteworthy that the optimal
covering and quantizer solutions are the ground states of many-body
interactions derived from $E_V(R)$
\cite{torquato_2010_reformulation, QuantizerRescale}.

The RSA procedure, which is the focus of the
present paper, is a time-dependent process to generate disordered hard-hypersphere packings in $\mathbb{R}^d$ 
\cite{torquato_2006_RSA, widom_1966_RSA, pomeau_1980_RSA, swendsen_1981_RSA, feder_1980_RSA, cooper_1987_RSA, tarjus_1991_RSA, renyi_1963_1d}. 
Starting with a large, empty region of $\mathbb{R}^d$ of volume $V$, spheres are randomly and sequentially placed 
into the volume subject to a nonoverlap constraint: 
if a new sphere does not overlap with any existing spheres, 
it will be added to the configuration; otherwise, the attempt is discarded.
One can stop the addition process at any time $t$, obtaining RSA configurations with various densities $\phi(t)$ 
up to the maximal saturation density $\phi_s=\phi(\infty)$ that occurs in the infinite-time limit. 
Besides identical d-dimensional spheres, the RSA packing process has also been investigated for 
polydisperse spheres \cite{adamczyk_1997_RSA_polydisperse, gray_2001_RSA_polydisperse} 
and other particle shapes, including squares \cite{brosilow1991random}, rectangles \cite{vigil1989random, vigil1990kinetics}, ellipses \cite{talbot1989unexpected, sherwood_1999_RSA_elipse}, spheroids \cite{sherwood_1999_RSA_elipsoid}, superdisks \cite{gromenko2009random}, sphere dimers \cite{ciesla2013modelling}, and sphere polymers \cite{ciesla2013random2} in $\mathbb{R}^d$, and for different shapes on lattices \cite{cadilhe2007random} and fractals \cite{ciesla2012random, ciesla2013random}. 
The RSA packing process in the first three space dimensions has been widely used to model the structure of cement paste \cite{xu_2013_RSA_application}, 
ion implantation in semiconductors \cite{roman_1983_RSA_application}, protein adsorption \cite{feder_1980_RSA_application}, 
polymer oxidation \cite{flory_1939_RSA_application}, and particles in cell membranes \cite{finegold_1979_RSA_application}.
The one-dimensional case, also known as the ``car-parking'' problem, 
has been solved analytically and its saturation density is $\phi=0.7475979202...$ \cite{renyi_1963_1d}.
However, for $d \ge 2$, the saturation density of RSA spheres has only been estimated through numerical simulations.

In general, generating exactly saturated (infinite-time limit) RSA configurations in $\mathbb{R}^d$ is particularly difficult because infinite computational time is not available. 
The long-time limit of RSA density behaves as \cite{feder_1980_RSA, pomeau_1980_RSA, swendsen_1981_RSA}:
\begin{equation}
\phi(\infty)-\phi(t) \sim t^{-1/d}.
\label{RSA_infinite_time_density}
\end{equation}
Previous investigators have
attempted to ascertain the saturation densities of RSA configurations by extrapolating the densities obtained at large, finite
times using the asymptotic formula (\ref{RSA_infinite_time_density}) \cite{cooper_1987_RSA, tarjus_1991_RSA, torquato_2006_RSA}.

In order to describe more efficient ways of generating nearly-saturated and fully-saturated RSA configurations, we first need to define two important concepts: the {\it exclusion sphere} and the {\it available space}. The {\it exclusion sphere} associated with a hard sphere of diameter $D$ (equal to $2R$) is the volume excluded to another hard sphere's center due to the impenetrability constraint, and thus an exclusion sphere of radius $D$ circumscribes a hard sphere. The {\it available space} is the space exterior to the union of the exclusion spheres of radius $D$ centered at each sphere in the packing. A more general notion of the available space is a fundamental ingredient in the formulation of a general canonical $n$-point distribution function \cite{torquato_1986_two_phase_media}.

An efficient algorithm to generate nearly-saturated RSA configurations was introduced in Ref.~\onlinecite{torquato_2006_RSA}. This procedure exploited an economical procedure to ascertain
the available space (as explained in the subsequent section).
Although a huge improvement in efficiency can be achieved, 
this and all other previous algorithms still require extrapolation of the density of nearly-saturated configurations to estimate the saturation limit.

In this paper, we present an improvement of the algorithm described in Ref.~\onlinecite{torquato_2006_RSA} in order to generate saturated (i.e., infinite-time limit) RSA packings of identical spheres in a finite amount of computational time. Using this algorithm, we improve upon previous calculations of the saturation packing and covering densities, pair correlation function, structure factor, void exclusion probability, and quantizer error in dimensions 2 through 8.

The rest of the paper is organized as follows: In Sec.~\ref{algorithm}, we describe the improved algorithm; in Sec.~\ref{results}, we present the packing and covering densities, pair correlation function, structure factor, void exclusion probability, and quantizer error of saturated RSA configurations; and in Sec.~\ref{conclusion}, we conclude with some discussions of extending this method to generate saturated RSA packings of objects other than congruent spheres.

\section{Improved Algorithm to Generate Saturated RSA Packings in $\mathbb{R}^d$}
\label{algorithm}

Reference \onlinecite{torquato_2006_RSA} introduced an efficient algorithm to generate nearly
saturated RSA configurations of hard $d$-dimensional spheres. 
Specifically, a hypercubic simulation box is divided into small hypercubic ``voxels'' 
with side lengths much smaller than the diameter of the spheres. At any instant of time, 
spheres are sequentially added to the simulation box whenever there is available
space for that sphere. Each voxel can be probed to determine whether it may contain any 
available space or not to add another sphere. By tracking all of the voxels that can 
contain some portion of the available space, one can make insertion attempts only 
inside these ``available voxels'' and save computational time. 
This enables one to achieve a huge improvement in computational efficiency
over previous methods. 
However, this and all other previous algorithms still require extrapolation of the density of nearly-saturated configurations to estimate the saturation limit.

\begin{figure}
\includegraphics[width=160mm]{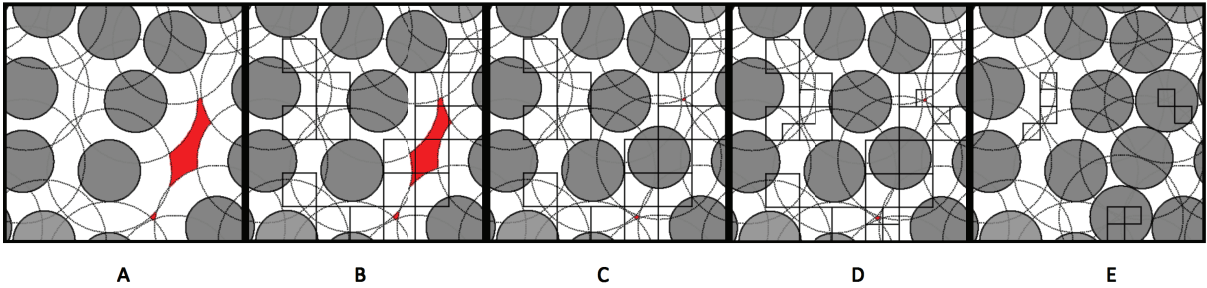}
\caption{A description of the key steps involved
to generate two-dimensional saturated RSA packings in a square box under periodic boundary conditions. Gray circles are RSA disks and dotted circles are their corresponding exclusion disks. The shaded region (red region in colored version) is the available space. Black squares are voxels in the available voxel list. A: Configuration after the first step. B: Same configuration with the available voxel list generated in the second step. C: A new disk is inserted in the third step, reducing the available space. D: In the fourth step, each available voxel is subdivided into $2^2$ sub-voxels. The available ones constitute a new voxel list. E: Return to the third step with the new available voxel list and two additional disks are inserted. The program then subdivides each voxel into four subvoxels and all subvoxels can be identified as unavailable. Thus the program finishes.}
\label{process}
\end{figure}

The improved algorithm reported in the present paper differs from the original voxel method \cite{torquato_2006_RSA} by dividing the undetermined voxels (voxels that are not included in any exclusion sphere after certain amount of insertion trials) into smaller subvoxels. Repeating this voxel subdivision process with
progressively greater resolution enables us to track the available space more and more precisely. Eventually, this allows us to discover {\it all} of the available space at any point in time and completely
consume it in order to arrive at saturated configurations. 

The improved algorithm consists of the following steps, which are illustrated in Figure~\ref{process}:
\begin{enumerate} 
\item Starting from an empty simulation box in $\mathbb R^d$, the Cartesian coordinates of a sphere of radius $R$ are randomly generated. This sphere is added if it does not overlap with
any existing sphere in the packing
at that point in time; otherwise,
the attempt is discarded. This addition process is repeated until the success rate is sufficiently low \cite{stoppingcriteria}. The acceptance ratio of this step equals to the volume fraction of the available space inside the simulation box:

\begin{equation}
P_{acceptance}=\Phi_{available}=\frac{V_{available}}{L^d},
\end{equation}

where $P_{acceptance}$ is the acceptance ratio of this step, $\Phi_{available}$ is the volume fraction of the available space, $V_{available}$ is the volume of the available space and $L^d$ is the volume of the simulation box with side length $L$.

\item When the fraction of the available space is low, we improve the acceptance ratio by avoiding insertion attempts in the unavailable space. To do this, the simulation box is divided into hypercubic voxels, with side lengths comparable to the sphere radius. Each voxel is probed to determine whether it is completely included in any of the exclusion spheres or not. If not, the voxel is added to the available voxel list. Thus we obtain an ``available voxel list''. A voxel in this list may or may not contain available space, but the voxels not included in this list are guaranteed to contain no available space.

\item Since some unavailable space is excluded from the voxel list, we can achieve a higher success rate of insertion by selecting a voxel randomly from the available voxel list, generate a random point inside it, attempt to insert a sphere and repeat this step. The acceptance ratio of this step is equal to the volume fraction of the available space inside voxels from the available voxel list:

\begin{equation}
P_{acceptance}=\Phi_{available}=\frac{V_{available}}{N_{voxel}V_{voxel}},
\end{equation}

where $P_{acceptance}$ is the acceptance ratio of this step, $\Phi_{available}$ is the volume fraction of the available space inside the voxel list, $V_{available}$ is the volume of the available space, $N_{voxel}$ is the number of voxels in the available voxel list and $V_{voxel}$ is the volume of a voxel.

\item In the previous step, spheres were inserted into the system, thus the volume of the available space will decrease. Eventually, $V_{available}$ is very low and $P_{acceptance}$ is also low. Thus we improve the efficiency again by dividing each voxel in the voxel list into $2^d$ sub-voxels, each with side length equal to a half of that of the original voxel. Each sub-voxel is checked for availability according to the rule described in step 2. The available ones constitute the new voxel list.

\item Return to step 3 with the new voxel list and repeat steps 3 to 5 until the number of voxels in the latest voxel list is zero. Since we only exclude a voxel from the voxel list when we are absolutely sure that it does not contain any available space, we know at this stage that the entire simulation box does not contain any available space and thus the configuration is saturated.
\end{enumerate}

\section{Results}
\label{results}
\subsection{Saturation Density}

We have used the method described in Sec.~\ref{algorithm} to generate saturated configurations of RSA packings
of hyperspheres in dimensions two
through eight in a hypercubic ($d$-dimensional cubic) box of side length $L$ under periodic boundary conditions. In each dimension, multiple sphere sizes are chosen. The relative sphere volume is represented by the ratio of a sphere's volume to the simulation box's volume $v_1(R)/L^d$, where $R$ is the sphere radius and $L^d$ is the volume of the hypercubic simulation box. For each sphere size, multiple configurations are generated. The number of spheres $N$ contained in these configurations fluctuate around some average value inversely proportional to $v_1(R)/L^d$. 
The relative sphere volume $v_1(R)/L^d$ and number of configurations $n_c$ generated for each sphere radius $R$ in each dimension is given in Table~\ref{NbrConfig}. The mean density and its standard error for each sphere radius $R$ is calculated. Subsequently, we plot the mean density $\phi_s$ and its standard error $\sigma$ versus a quantity proportional to $N^{-1/2}$, namely $[v_1(R)/L^d]^{1/2}$. We then perform a weighted linear least squares fit \cite{strutz2010data} to this function in each dimension in order to extrapolate to the infinite-system-size [$v_1(R)/L^d\to 0$] limit. The weight is given by
\begin{equation}
W(R)=\frac{1}{\sigma^2(R)},
\end{equation}
where $\sigma(R)$ is the standard error of the mean density for spheres with radius $R$.

\begin{table}[h!]
\setlength{\tabcolsep}{12pt}
\caption{Dimensionless sphere size $v_1(R)/L^d$ and number of configurations $n_c$ generated for each dimension $d$.}
\begin{tabular}{c | c c c c c}
\hline
\multirow{2}{*}{$d=2$} & $v_1(R)/L^d$ & $1.0884\times 10^{-7}$ & $5.4420\times 10^{-8}$ & $2.7210\times 10^{-8}$ & $1.3605\times 10^{-8}$\\
& $n_c$ & $250$ & $250$ & $250$ & $250$ \\ \hline

\multirow{4}{*}{$d=3$} & $v_1(R)/L^d$ & $3.82925\times 10^{-7}$ & $1.91462\times 10^{-7}$ & $7.65850\times 10^{-8}$ & $3.82925\times 10^{-8}$\\
& $n_c$ & $250$ & $250$ & $250$ & $250$ \\ \cline{2-6}

& $v_1(R)/L^d$ & $1.91462\times 10^{-8}$ & & & \\
& $n_c$ & $250$ & & & \\ \hline

\multirow{4}{*}{$d=4$} & $v_1(R)/L^d$ & $5.20225\times 10^{-6}$ & $2.60112\times 10^{-6}$ & $1.30056\times 10^{-6}$ & $5.20225\times 10^{-7}$\\
& $n_c$ & $250$ & $250$ & $250$ & $250$ \\ \cline{2-6}

& $v_1(R)/L^d$ & $2.60112\times 10^{-7}$ & $1.30056\times 10^{-7}$ & & \\
& $n_c$ & $250$ & $250$  & & \\ \hline

\multirow{4}{*}{$d=5$} & $v_1(R)/L^d$ & $1.71000\times 10^{-5}$ & $8.55000\times 10^{-6}$ & $3.42000\times 10^{-6}$ & $1.71000\times 10^{-6}$\\
& $n_c$ & $250$ & $250$ & $250$ & $250$ \\ \cline{2-6}

& $v_1(R)/L^d$ & $8.55000\times 10^{-7}$ & $3.42000\times 10^{-7}$ & & \\
& $n_c$ & $250$ & $250$  & & \\ \hline

\multirow{4}{*}{$d=6$} & $v_1(R)/L^d$ & $2.22500\times 10^{-5}$ & $1.11250\times 10^{-5}$ & $5.56250\times 10^{-6}$ & $2.78125\times 10^{-6}$\\
& $n_c$ & $50$ & $50$ & $50$ & $50$ \\ \cline{2-6}

& $v_1(R)/L^d$ & $1.39062\times 10^{-6}$ & & & \\
& $n_c$ & $50$ & & & \\ \hline

\multirow{4}{*}{$d=7$} & $v_1(R)/L^d$ & $2.72744\times 10^{-5}$ & $1.36372\times 10^{-5}$ & $6.81859\times 10^{-6}$ & $4.54573\times 10^{-6}$\\
& $n_c$ & $70$ & $30$ & $20$ & $20$ \\ \cline{2-6}

& $v_1(R)/L^d$ & $3.40930\times 10^{-6}$ & $1.94817\times 10^{-6}$ & $1.36372\times 10^{-6}$ & \\
& $n_c$ & $20$ & $20$ & $15$ & \\ \hline

\multirow{2}{*}{$d=8$} & $v_1(R)/L^d$ & $4.16930\times 10^{-5}$ & $2.08465\times 10^{-5}$ & $1.38977\times 10^{-5}$ &\\
& $n_c$ & $11$ & $7$ & $5$ &\\ \hline
\end{tabular}
\label{NbrConfig}
\end{table}

\begin{figure}[H]
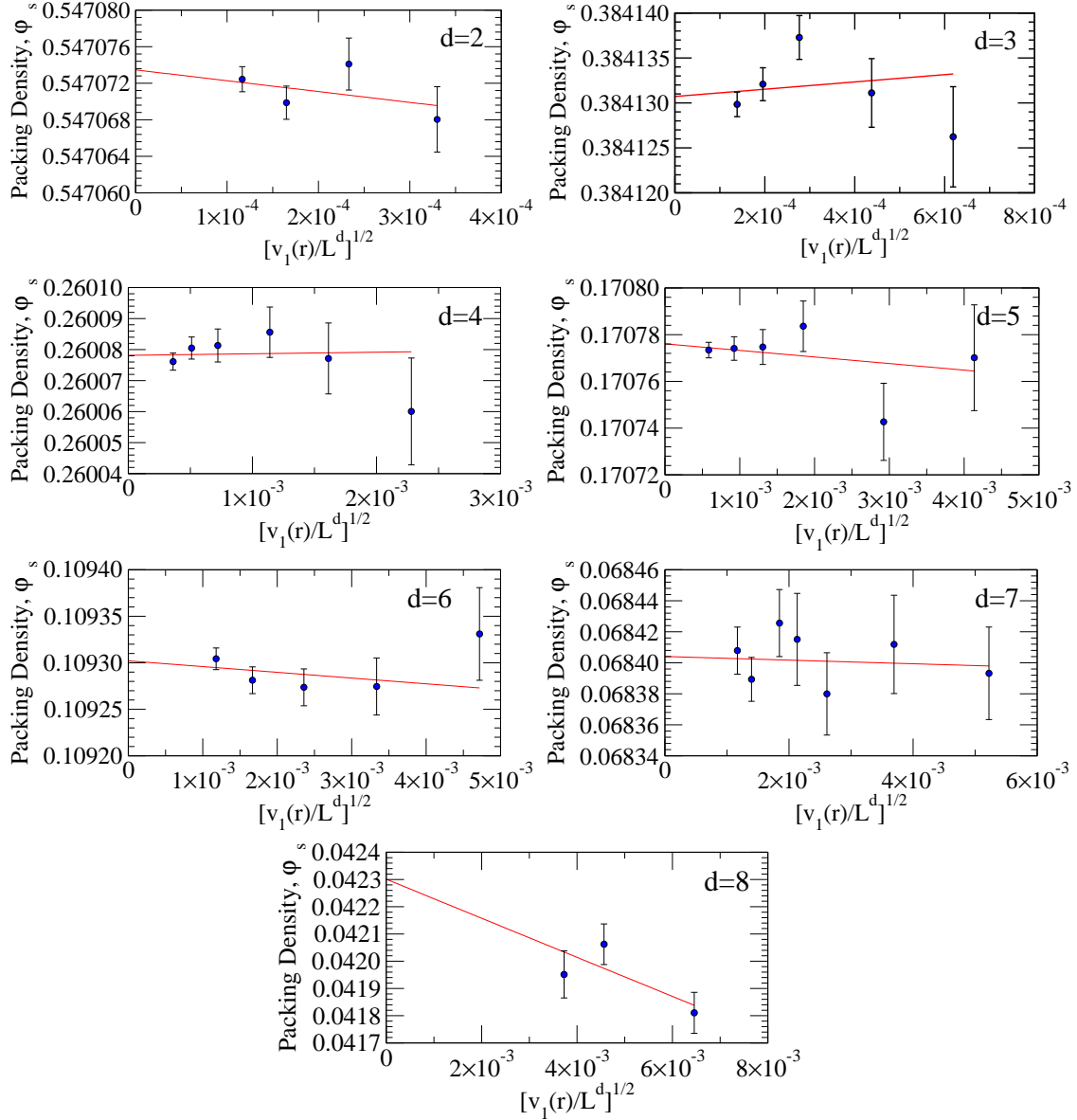

\centering
                 \includegraphics[width=0.45\textwidth]{Fig2a.eps}
                 \includegraphics[width=0.45\textwidth]{Fig2b.eps}
                 \includegraphics[width=0.45\textwidth]{Fig2c.eps}
                 \includegraphics[width=0.45\textwidth]{Fig2d.eps}
                 \includegraphics[width=0.45\textwidth]{Fig2e.eps}
                 \includegraphics[width=0.45\textwidth]{Fig2f.eps}
                 \includegraphics[width=0.45\textwidth]{Fig2g.eps}
\caption{RSA saturation packing density, $\phi_s$, (filled circles) of different system sizes as measured by a quantity proportional to $N^{-1/2}$, namely $[v_1(R)/L^d]^{1/2}$, in different dimensions $d$. Included are the associated linear fits. Error bars associated with filled circles are the standard error of the mean as obtained from averaging multiple configurations.}
\label{density}
\end{figure}

The mean densities and the associated standard errors for different sphere radii $R$ are shown in Fig.~\ref{density}, while the extrapolated infinite-system-size densities are shown in Table~\ref{infdensity}.
These density estimates for $2 \le d \le 8$ have been determined with heretofore unattained accuracy, including in the most previously studied dimensions of $d=2$ and $d=3$.
For $d=2$, several previous studies
produced the following density estimates 
$0.547\pm0.002$ \cite{feder_1980_RSA}, 
$0.547\pm0.003$ \cite{hinrichsen1986RSA}, 
and $0.54700\pm0.000063$ \cite{torquato_2006_RSA}.
For $d=3$, several previous investigations
yielded the following density estimates 
$0.37-0.40$ \cite{cooper_1987_RSA},
$0.385\pm0.010$ \cite{cooper1988RSA},
$0.382\pm0.0005$ \cite{talbot1991RSA},
and $0.38278\pm0.000046$ \cite{torquato_2006_RSA}.
Compared with previous results of saturation densities for $2 \le d \le 6$ \cite{torquato_2006_RSA}, our corresponding results are only slightly higher for two dimensions, but the discrepancy increases as dimension increases. 
This suggests that the previous attempts did not generate fully saturated configurations, especially in high dimensions. Table \ref{infdensity} also includes corresponding RSA covering densities. 
A RSA covering is obtained by replacing each sphere in a saturated RSA packing in $\mathbb{R}^d$ with its exclusion sphere, and thus its covering density is given by
\begin{equation}
\theta=2^d\phi_s.
\end{equation}

\begin{table}
\setlength{\tabcolsep}{12pt}
\caption{RSA saturation densities and covering densities in different dimensions, extrapolated to the infinite system size limit. Here $\phi_s$ is saturation packing density and $\theta$ is the corresponding covering density.}
\begin{tabular}{c c c c}
\hline
Dimension & $\phi_s$ [Present Work] & $\phi_s$ [Ref.~\onlinecite{torquato_2006_RSA}] & $\theta$ [Present Work] \\ \hline
2 & $0.5470735\pm0.0000028$ & $0.54700\pm0.000063$ & $2.188294\pm0.000011$\\
3 & $0.3841307\pm0.0000021$ & $0.38278\pm0.000046$ & $3.073046\pm0.000017$\\
4 & $0.2600781\pm0.0000037$ & $0.25454\pm0.000091$ & $4.161250\pm0.000060$\\
5 & $0.1707761\pm0.0000046$ & $0.16102\pm0.000036$ & $5.46483\pm0.00015$\\
6 & $0.109302\pm0.000019$ & $0.09394\pm0.000048$ & $6.9953\pm0.00012$\\
7 & $0.068404\pm0.000016$ & $$ & $8.75572\pm 0.0020$\\
8 & $0.04230\pm0.00021$ & $$ & $10.829\pm0.053$\\
\hline
\end{tabular}
\label{infdensity}
\end{table}

\subsection{Pair Correlation Function and Structure Factor}
We have used the methods described in Ref.~\onlinecite{torquato_2006_RSA} to calculate the pair correlation function $g_2(r)$ and structure factor $S(k)$ of the saturated RSA configurations for $2 \le d \le 7$. [For $d=8$, we can only generate relatively small configurations, which are not suitable to calculate $g_2(r)$ and $S(k)$ accurately.] The structure factor is calculated using the collective density variables approach, i.e.,
\begin{equation}
S(\mathbf k)=\frac{\langle| \tilde{\rho}(\mathbf k)^2|\rangle}{N}, 
\end{equation}
where $N$ is the number of spheres in the periodic hypercubic box of side length $L$,
\begin{equation}
\tilde{\rho}(\mathbf k) = \sum_{j=1}^N \exp(i \mathbf k \cdot \mathbf r_j)
\end{equation}
is the complex collective density
variable and
\begin{equation}
\mathbf k = (\frac{2 \pi n_1}{L}, \frac{2 \pi n_2}{L}, ... , \frac{2 \pi n_d}{L}),
\end{equation}
where $\mathbf k$ is a wave vector and where $n_i$ ($i=1,2,\ldots,d$) are
the integers. In presenting the structure factor, we will omit the forward
scattering contribution ($\bf k=0$); see Ref.~\onlinecite{torquato_2006_RSA} for additional details.

\begin{figure}
\begin{center}
\includegraphics[width=150mm]{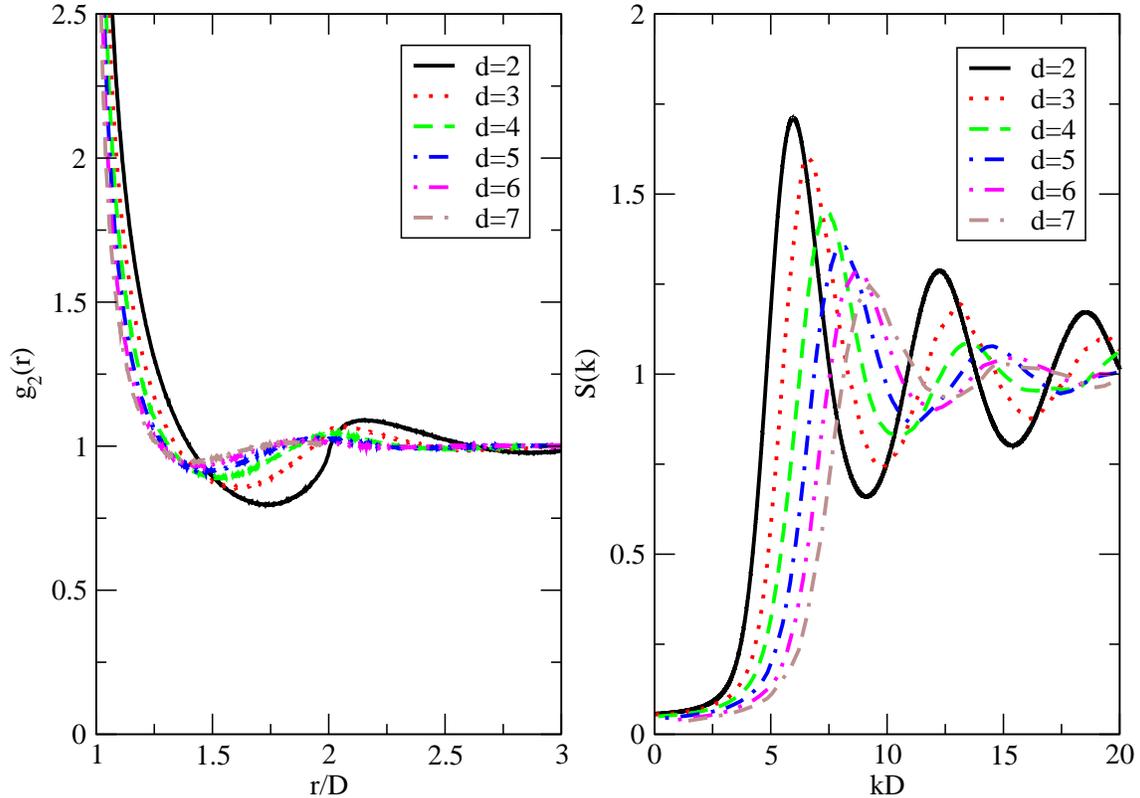}
\end{center}
\caption{Pair correlation function and structure factor of saturated RSA configurations, in two through seven dimensions.  It is clearly seen that
these pair statistics indicate that the packings become more decorrelated
as the dimension increases.}
\label{Sg}
\end{figure}

These pair statistics are shown in Figure~\ref{Sg} for dimensions
two through seven. The decorrelation exhibited with increasing dimension was also observed in Ref.~\onlinecite{torquato_2006_RSA}.
These trends are clearly consistent with a recently proposed ``decorrelation'' principle, which states that {\it unconstrained} spatial correlations diminish as the dimension increases and vanish in the $d \to \infty$ limit \cite{torquato_2006_ghostRSA, torquato_2006_lower_bound_packing}.
It is noteworthy that decorrelation
is already exhibited in these low dimensions,
which has been observed for other
types of hard-sphere packings \cite{torquato_2006_ghostRSA, skoge2006packing}.

The pair correlation function $g_2(r)$ of saturated RSA configurations has a logarithmic singularity when $r$ approaches the sphere diameter, $D$ \cite{pomeau_1980_RSA, swendsen_1981_RSA} :
\begin{equation}
g_2(r) \sim -\ln(r/D-1), \mbox{ } r \to D^+.
\end{equation}
Based on this analytical form, we have fit our pair correlation functions at $D<r<1.018D$ to the following formula:
\begin{equation}
g_2(r) =a_0\ln(r/D-1)+a_1.
\end{equation}
Our results are shown in Table~\ref{g2_near}. The absolute value of $a_0$ in each dimension are significantly higher than previous results \cite{torquato_2006_RSA}, which means that our $g_2(r)$'s are much sharper near $r=D$. 
This is due to the fact that our algorithm is capable of finding even the smallest fragments of the available space. Finding those pieces enables us to insert spheres that are very close to other spheres, substantially increasing $g_2(r)$ near $r=D$.

It is of interest to see to what extent
RSA packings are hyperuniform. A packing is {\it hyperuniform} if the structure factor in the zero-wavenumber limit, $S_0 \equiv \lim_{k \to 0} S(k)$, is zero \cite{torquato_2003_hyperuniform, zachary2009hyperuniformity}. Thus, the magnitude of $S_0$ quantifies the ``distance'' from hyperuniformity. It was reported in Ref.~\onlinecite{torquato_2006_RSA} that $S_0$ of saturated RSA packings decreases with dimension but because these
simulations were not as precise
in higher dimensions, the high-$d$ asymptotic behavior of $S_0$
was difficult to ascertain. We fit the structure factors that we have determined in the present paper to a function of the form $S(k)=S_0+S_2 k^2+S_4 k^4$ in each dimension near $k = 0$ in order to estimate $S_0$. This form is the exact
behavior of the structure factor
as $k$ goes to zero, as shown in Ref.~\onlinecite{torquato_2006_RSA}. The results for $S_0$ are summarized in Table~\ref{S0}. It is seen that as $d$ increases,
$S_0$ decreases, i.e., the ``degree of hyperuniformity'' (the ability
to suppress infinite-wavelength
density fluctuations) increases.
The data indicates that $S_0$
tends to the perfect hyperuniformity
limit of zero as $d \rightarrow \infty$.
As we will show in Sec.~\ref{Sec_Ev_G}, in the $d \rightarrow \infty$ limit, the void exclusion probability of RSA packings tends to a step function \cite{torquato_2010_reformulation}. 
This indicates that the vacancies in infinite-dimensional RSA packings are spherically-shaped with similar sizes.
Thus, $S_0$ tends to zero in the $d \rightarrow \infty$ limit.
This also explains why RSA packings become more stealthy [$S(k)$ is nearly zero for larger range of $k$ near $k=0$] \cite{batten2011inherent} as $d$ increases.

\begin{table}
\setlength{\tabcolsep}{12pt}\caption{Results from fitting data to $g_2(r) =a_0\ln(r/D-1)+a_1$ in the near-contact range $D<r<1.018D$}
\begin{tabular}{c c c}
\hline
Dimension & $a_0$ & $a_1$ \\ \hline
2 & $-1.562\pm0.031$ & $-2.155\pm0.155$\\
3 & $-1.603\pm0.026$ & $-2.709\pm0.133$\\
4 & $-1.488\pm0.028$ & $-2.582\pm0.116$\\
5 & $-1.396\pm0.030$ & $-2.565\pm0.155$\\
6 & $-1.200\pm0.039$ & $-1.984\pm0.206$\\
7 & $-1.169\pm0.055$ & $-2.116\pm0.269$\\
\hline
\end{tabular}
\label{g2_near}
\end{table}

\begin{table}
\caption{Structure factor $S(k)$ at $k=0$, obtained by fitting data to $S(k)=S_0+S_2 k^2+S_4 k^4$ at $0<kD<3$, where $S_0$, $S_2$, and $S_4$ are fitting parameters.}
\begin{tabular}{c c}
\hline
Dimension & $S_0$\\ \hline
2 & $0.05869\pm0.00004$\\
3 & $0.05581\pm0.00005$\\
4 & $0.05082\pm0.00007$\\
5 & $0.04544\pm0.00029$\\
6 & $0.03834\pm0.00072$\\
7 & $0.03140\pm0.00173$\\
\hline
\end{tabular}
\label{S0}
\end{table}

\subsection{Void Exclusion Probability and Quantizer Error}
\label{Sec_Ev_G}
\begin{figure}
\begin{center}
\includegraphics[width=150mm]{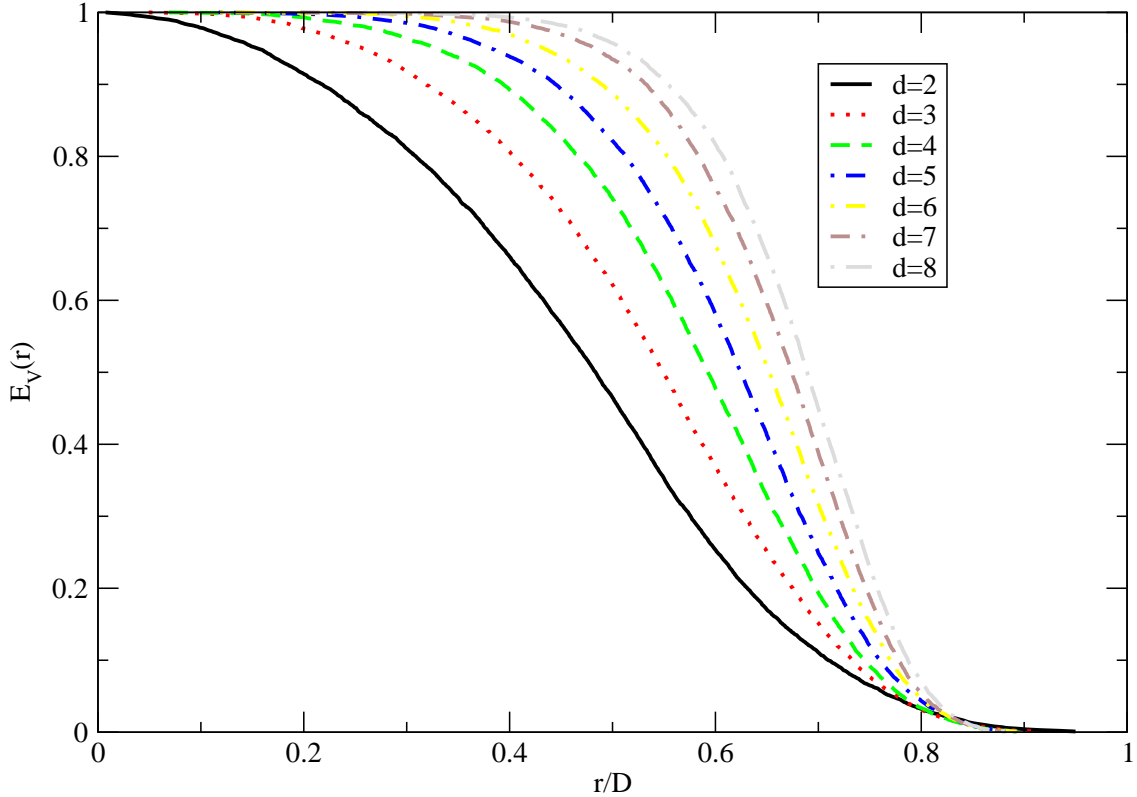}
\end{center}
\caption{Void exclusion probability of saturated RSA configurations, in two through eight dimensions.}
\label{Ev}
\end{figure}

We have calculated the void exclusion probability $E_V(r)$ (discussed in the Introduction) of saturated RSA configurations for $2 \le d \le 8$ and findings are summarized in Figure~\ref{Ev}. 
The void exclusion probability in all dimensions vanishes at $r \to D^-$, confirming that the exclusion spheres with radius $R_c=D$ cover the space and that our RSA configurations are saturated.
Our results are similar to previously reported results \cite{torquato_2010_reformulation} and strongly supports the theory that the void exclusion probability of RSA packings tend to a step function in the infinite-dimensional limit \cite{torquato_2010_reformulation}, i.e.,
\begin{equation}
E_V(r) \to \Theta(r-D) \mbox{\hspace{1cm} } (d \to \infty),
\end{equation}
where
\begin{equation}
\Theta(x) = \left\{ \begin{matrix} 0,\mbox{\hspace{1cm} } x < 0 \\
 1,\mbox{\hspace{1cm} } x \ge 0 \end{matrix} \right.
\end{equation}
is the Heaviside step function.
This indicates that the ``holes'' in RSA packings become spherically-shaped with similar sizes as $d$ tends to infinity.
It is interesting to note that the void exclusion probability of fermionic systems have similar behavior in the high-dimensional limit \cite{torquato2008point}.

We have calculated the quantizer error $\mathcal G$ for saturated RSA configurations
for $2 \le d \le 8$. These results are summarized in Table~\ref{quantizer_error}. Compared with results in Ref.~\onlinecite{torquato_2010_reformulation} for $2 \le d \le 6$, our corresponding results for $\mathcal G$ are somewhat lower.

\begin{table}
\setlength{\tabcolsep}{6pt}
\caption{Scaled dimensionless quantizer error $\mathcal{G}$.}
\begin{tabular}{c c c}
\hline
Dimension & $\mathcal{G}$ [Present Work] & $\mathcal{G}$ [Ref.~\onlinecite{torquato_2010_reformulation}] \\ \hline
2 & $0.08848\pm0.00018$ & $0.09900$\\
3 & $0.08441\pm0.00013$ & $0.09232$\\
4 & $0.08154\pm0.00011$ & $0.08410$\\
5 & $0.07936\pm0.00009$ & $0.07960$\\
6 & $0.07765\pm0.00007$ & $0.07799$\\
7 & $0.07623\pm0.00007$ & \\
8 & $0.07508\pm0.00009$ & \\
\hline
\end{tabular}
\label{quantizer_error}
\end{table}

\section{Conclusions and Discussion}
\label{conclusion}
We have devised an efficient algorithm to generate exactly saturated, infinite-time limit RSA configurations in finite computational time across Euclidean space dimensions. With the algorithm, we have improved previous results of the saturation density and extended them to a wider
range of dimensions, i.e., up through
dimension eight. The associated covering density, pair correlation function, structure factor, void exclusion probability, and quantizer error have also been improved. In particular, we found appreciable improvement for $g_2(r)$ near contact and $S(k)$ in the $k \to 0$ limit, which are especially sensitive to whether or not very small fragments
of the available space are truly eliminated as the saturation state is approached.
We observed that as $d$ increases, the degree of ``hyperuniformity"
(the magnitude of the suppression of infinite-wavelength
density fluctuations) increases and appears to be consistent with $\lim_{d \to \infty} S(0) = 0.$ Our results also supports the ``decorrelation principle'', which in turns lends further credence to a conjectural lower bound on the maximal sphere packing density that provides the putative exponential improvement on Minkowski's lower bound \cite{torquato_2006_lower_bound_packing}.

It is noteworthy that the RSA packing
in $\mathbb{R}^d$ has relevance
in the study of high-dimensional scaling of packing densities. For example, Ref.~\onlinecite{torquato_2006_RSA} suggested that since RSA packing densities appear to have a similar scaling in high dimensions as the best lower bound on Bravais lattice packings densities, the density of disordered packings might eventually surpass that of the densest lattice packing beyond some large
but finite dimension. Our improvements to the saturation densities, as well as a previous investigation \cite{torquato_2006_lower_bound_packing}, support this conjecture. 
Converting a packing into a covering by replacing each sphere with its exclusion sphere is rigorous only if the packing is exactly saturated. By guaranteeing that the packings that we generated are saturated, we rigorously met this condition (in a large finite simulation box). Although the best known lattice covering and lattice quantizer
perform better than their RSA counterparts in low dimensions, RSA packings may outperform lattices in sufficiently high dimensions, as suggested in Ref.~\onlinecite{torquato_2010_reformulation}.

It is useful here to comment on the ability to ascertain the high-dimensional scaling of RSA packing densities from low-dimensional data \cite{torquato_2010_reformulation, torquato_2006_RSA}.
We have fitted our data of the saturation densities as a function of $d$ for $2 \le d \le 8$ using a variety of different functions. The best fit we find is the following form:
\begin{equation}
\label{fit2}
\phi_s=\frac{a_1+a_2d+a_3d^2}{2^d},
\end{equation}
where $a_1=1.0801$, $a_2=0.32565$, and $a_3=0.11056$ are parameters.
However, it is not clear how accurate this form is for $d \ge 9$.
In fact, this form is likely not correct in high dimensions, where it has been suggested from theoretical considerations \cite{torquato_2010_reformulation} that high-dimensional scaling may be given by the asymptotic form
\begin{equation}
\label{fit1}
\phi_s=\frac{b_1+b_2d+b_3d\ln(d)}{2^d},
\end{equation}
where $b_1$, $b_2$, and $b_3$ are constants.
It is noteworthy that (\ref{fit1}) provides
a fit that is very nearly as good as (\ref{fit2}).
Nonetheless, for $d=15$, the estimates
of the saturation densities  obtained
from (\ref{fit2}) and (\ref{fit1}) differ by about 20\%,
which is a substantial discrepancy
and indicates the uncertainties involved
in applying such dimensional scalings for even moderately-sized dimensions. When $d$ is very large, extrapolations based
on fits of low-dimensional data
is even more problematic.
In this limit, Eq.~(\ref{fit2}) is dominated by the $a_3d^2/2^d$ term, which can be significantly larger than the $a_3d\ln(d)/2^d$ dominating term in Eq.~(\ref{fit1}), although it is safe to say that the saturation density grows at least as fast as $d 2^{-d}$.
Therefore, caution should be exercised
in attempting to ascertain the precise high-$d$
asymptotic behavior of RSA saturation densities from our data in relatively low dimensions.
The same level of caution should be employed in attempting to determine high-$d$ scaling behavior by extrapolating low-dimensional packing densities for other types of sphere packings.
For example, it may useful to revisit the high-dimensional
scalings that have been ascertained or tested for the maximally random jammed densities
 \cite{parisi2010mean, charbonneau2011glass}. 
In summary, it is nontrivial to ascertain high-$d$ scalings of packing densities from low-dimensional information.
In contrast, in the study of the dimensional dependence of continuum percolation thresholds, it is possible to obtain exact high-$d$ asymptotics and tight upper and lower bounds that apply across all dimensions \cite{torquato2012effect, torquato2012effect2}.

RSA packings of spheres with a polydispersity in size have also been investigated previously \cite{adamczyk_1997_RSA_polydisperse, gray_2001_RSA_polydisperse}. 
Our algorithm can easily be extended to generate saturated RSA packings of polydisperse spheres in $\mathbb R^d$ 
by constructing a $(d+1)$-dimensional auxiliary space for the associated radius-dependent available space and voxels, 
where the additional dimension is used to represent the radius of a sphere
that could be added in the RSA process.
RSA packings of nonspherical particles have also been studied, including squares \cite{brosilow1991random}, rectangles \cite{vigil1989random, vigil1990kinetics}, ellipses \cite{talbot1989unexpected, sherwood_1999_RSA_elipse}, spheroids \cite{sherwood_1999_RSA_elipsoid}, and superdisks \cite{gromenko2009random}.
While packings of polyhedra
have received recent attention \cite{chan1991effective, torquato2012organizing}, RSA packings of such
shapes have not been considered
to our knowledge.
Our algorithm can also be extended to treat these situations by constructing auxiliary spaces for the associated orientation-dependent available space and voxels. 
The dimension of such an auxiliary space is determined by the total number of degrees of freedom associated with a particle, i.e., translational and rotational degrees of freedom. 
The extensions of the methods devised here to generate saturated packings of polydisperse spheres and nonspherical particles is an interesting direction for future research.

\begin{acknowledgments}
We are very grateful to {\'E}tienne Marcotte, Yang Jiao, and Adam Hopkins for many helpful discussions and Yang Jiao, Steven Atkinson, and {\'E}tienne Marcotte for their careful reading of the manuscript.
This work was supported by the Materials Research
Science and Engineering Center Program of the National
Science Foundation under Grant No. DMR-0820341 and
by the Division of Mathematical Sciences at the National
Science Foundation under Award No. DMS-1211087. This
work was partially supported by a grant from the Simons
Foundation (Grant No. 231015 to Salvatore Torquato).
\end{acknowledgments}


\begin{thebibliography}{10}
\bibitem{stoppingcriteria}In the implementation
of our algorithm, the criteria for ``sufficiently low'' is less than 3 spheres inserted in $N$ trials. The optimal choice of $N$ depends on the dimension $d$, ranging from $N=500$ for $d=2$ to $N=2 \times 10^8$ for $d=8$.
\end{thebibliography}
\end{document}